\documentstyle[12pt]{article}

\advance\textheight by -5cm 
\advance\textheight by -2.5cm 
\begin{document}
\begin{titlepage}
\begin{center}
October 4, 1993     \hfill    LBL-34703\\
\vskip .5in

{\large \bf Vacuum Misalignment in High Energy Collisions}
\footnote{This work was supported by the Natuaral Sciences and
Engineering Research Coucil of Canada and
by the Director, Office of Energy
Research, Office of High Energy and Nuclear Physics, Division of High
Energy Physics of the U.S. Department of Energy under Contract
DE-AC03-76SF00098.}

\vskip .5in
Zheng Huang\footnote{Electronic Address: huang@theorm.lbl.gov}\\
{\em Theoretical Physics Group\\
    Lawrence Berkeley Laboratory\\
      University of California\\
    Berkeley, California 94720}
\end{center}

\vskip .5in

\begin{abstract}
We study a recent proposal to observe the disoriented chiral condensate in high
energy collisions. In order to produce a large fluctuation in pion probability
distribution, a large size of the correlated region is essential. We study the
role of the intrinsic symmetry breaking and suggest that a negative
$(\mbox{mass})^2$ of mesons arising from a misaligned vacuum may be a candidate
for such a mechanism. We further point out a possibility to observe
unsuppressed
 strong $CP$ violation in the disoriented vacuum phase.
\end{abstract}
\end{titlepage}

\newpage
\renewcommand{\thepage}{\arabic{page}}
\setcounter{page}{2}
Recently it has been speculated that disoriented chiral condensates may be
observed in very high energy hadronic collisions where the temperatures are
high
and the energy density can be up to several $\mbox{GeV}/\mbox{fm}^3$
\cite{anselm,taylor}. It has been proposed to study this phenomenon at the
Tevatron
and at the SSC and the proposal has been approved for installation at the
Tevatron. The
experiment results are expected at the end of 1993. It is further argued that a
disoriented chiral condensate (DCC) may be a possible explanation of the
Centauro cosmic ray experiments \cite{centauro},
 where a large fluctuation in the
neutral pion distribution has been observed. In this paper, we study some
theoretical aspects of this scenario,  in particular, the role of intrinsic
symmetry breaking (ISB). When the vacuum is misaligned with the preferential
orientation determined by the ISB, the excited mesons are not stable and may
have
a negative $(\mbox{mass})^2$. We further point out a possibility
of observing unsuppressed $CP$ violation in the DCC phase.

In high energy collisions, the collision debris form a high temperature ($T$)
and high density environment. These debris are highly relativistic and will
expand outward at the speed of light, leaving behind a cool interior where the
temperature suddenly drops from $T$ to 0. If initially $T>T_c$
($T_c$ is the critical temperature of the chiral phase transition), the
vacuum orientation in the interior
may, in general, be different from that in the exterior, since the interior
is separated from
the exterior by a hot shell of debris. The
misaligned vacuum regions would relax to the true ground state (the state of
the empty exterior) by coherent pion emissions. If a coherent configuration of
pions are radiated along some particular direction in the isospin vector space,
the probability of finding a fixed ratio ($r$) of neutral pions to charged
pions
is given by the probability of finding an arbitrary isovector whose angle with
$\pi ^0$-direction is $\beta$ ($\cos ^2\beta =r$)
\begin{eqnarray}
dP(r)=\frac{1}{2} \frac{1}{\sqrt{r}} dr
\end{eqnarray}
where
\begin{eqnarray}
r=\frac{n_{\pi ^0}}{n_{\pi ^0}+n_{\pi ^{\pm}}}\; .
\end{eqnarray}
This probability distribution is very different from an ordinary Gaussian
distribution where one expects it to be sharply peaked about $1/3$.

There are many theoretical uncertainties in this picture, among which are the
role of the intrinsic symmetry breaking, that is, the quark current masses, and
the size of the correlated regions. In the linear $\sigma$-model,
\begin{eqnarray}
{\cal L}=\frac{1}{2}\partial_\mu\phi^\dagger\partial^\mu\phi
 -\frac{\lambda}{4}(\phi^{\dagger}\phi -v^2)^2+\frac{1}{2}
c(\phi ^\dagger +\phi)
\end{eqnarray}
where $\phi$ is a four component vector $(\sigma ,\vec{\pi})$ in O(4) or
equivalently $\phi =\sigma +i\vec{\pi}\cdot\vec{\tau}$, there is an explicit
symmetry breaking term $c\sigma$. Even at  high temperature, $c\sigma$ term
cannot be dropped since the temperature dependence of $c$ is weak. The least
free energy configuration at $T$ is not $(0,0)$ but
 $(\langle \sigma\rangle ,0)$.
As a result, there is a pre-existent preferential direction when the chiral
phase transition occurs. In principle, all possible oriented configurations
from
the ensemble must be {\em a prior} weighted by their corresponding free
energies
at temperature $T$. However, when $T\geq T_c$, the vacuum expectation value
(VEV) is very small ($\propto \frac{c^{1/3}}{\lambda}$) compared to $T$, the
 vacuum configurations with different orientations at the critical temperature
 do not have enough free energy to ``relax'' to the preferential direction if
 the system is out of equilibrium and the temperature drops fast enough.
Practically, these
 field configurations $(\sigma ,\vec{\pi})$ will have approximately
 equal probability, evolving toward the subcritical region in the
 manner of a quench.

The crucial question in this picture is the typical size of the correlated
region. If the dynamical evolution of the system is a process of cooling
through
equilibrium, one expects that domains in the interior will have different
vacuum
orientations, which individually evolve according to dynamics. Even if the
misalignment does occur independently in small regions, their effects, however,
are modest and the large fluctuations in the pion distribution will be washed
out by averaging the orientations. However, as pointed out by Rajagopal and
 Wilczek (RW) \cite{rw}, the situation can be qualitatively different if the
 cooling process is very rapid and the system is out of equilibrium. An analog
is
 quenching magnet by cooling a piece of hot iron. Typically what happens is the
 mismatch of the configurations and their evolution. After a quench, the high
 temperature configuration does not have time to ``react'' to the sudden change
 of the environment; instead, it will satisfy a zero temperature equation of
 motion. For example, the VEV's of $\phi$ would stay what it was at high
 temperature for a while, that is, $|\langle \phi \rangle|\sim 0$. But because
 the temperature is zero after the quench, the Goldstone boson
$(\mbox{mass})^2$ becomes
 negative
\begin{eqnarray}
\tilde{m}_\pi ^2=-\mu^2+2\lambda\langle \phi\rangle ^2\sim -\mu^2
\end{eqnarray}
and the long wavelength modes will grow with time and the short wavelength will
be suppressed. As a result, there will be a large size of the correlated region
in which the vacuum orientation may be unique, and the pion emission from this
region will be quite structural. This picture is also supported by some recent
studies on this issue in lower dimensions and a four-fermion interaction model
\cite{kogan}.

The appearance of negative $(\mbox{mass})^2$ modes due to the hesitation of the
VEV's of $\phi$
when the temperature suddenly changes offers a serious theoretical challenge in
the studies of the critical phenomena. Whether such a mechanism is also to
provide enough energy density for the pion emission is however unclear. When
$|\langle \phi\rangle |$ stays around zero, even though the long range
correlation may occur, the energy density of such a structure is small since it
is proportional to
$|\langle \phi\rangle |$. It is not clear that there will be enough pion
radiations to be observed before
$|\langle \phi\rangle |$ oscillates to its zero temperature value. When
$|\langle \phi\rangle |$ gets close to the zero temperature VEV's, this
mechanism cease to work and the correlated region would shrink into small
pieces.

However, we argue that another mechanism may be in effect to hold the
correlated
region together even longer, and in the mean time to provide a large energy
density. That is, when
$|\langle \phi\rangle |$ gets to its final value, the vacuum is still
misaligned
with the intrinsic symmetry breaking (ISB)
 since it takes longer for the vacuum orientation to relax due
to the small free energy difference; and the Goldstone bosons are not stable:
their $(\mbox{mass})^2$ can be negative if the misalignment
is big enough. The picture based
on this mechanism is in essence closer to  Ref.\ \cite{anselm} where the
amplitude of classical pion fields is equal to the zero temperature
VEV of $\phi$. A theoretical
idealization can be made by considering a zero temperature lagrangian while the
vacuum orientation is however pre-determined at high temperature (not by the
ISB).
The treatment is opposite to what we do in the perturbation theory
with degeneracy or in the vacuum alignment procedure \cite{dashen}. We have a
symmetric potential $V_0$ and a symmetry breaking $V_1$. In the perturbation
theory, one has to first compute the vacuum orientation {\em according} to
$V_1$
and expand $V_1$ in that basis. In our consideration, however, the
vacuum orientation is set
before the quench, and can in general mismatch with $V_1$. For example, one may
specify the VEV
\begin{eqnarray}
\left\langle \phi \right \rangle _{\theta ,\vec{n}}\approx (F_\pi \cos\theta,
F_\pi\sin \theta \vec{n})\equiv F_\pi e^{i\theta (\vec{n}\cdot \vec{\tau})}
\end{eqnarray}
where $\vec{n}$ is a unit vector in isospin space. The energy density of the
vacuum is determined by the perturbation $V_1$ in the lowest order
\begin{eqnarray}
E_\theta =-cF_\pi\cos\theta =-m_\pi^2F_\pi ^2\cos\theta \; .\label{a}
\end{eqnarray}
Since $\left . \frac{\partial E}{\partial \theta}\right |_{\theta \neq 0}\neq
0$, the vacuum is misaligned. Note that $E_\theta$ is larger than $E_0$ in the
exterior and the difference is $\Delta E_\theta =E_\theta -E_0=
-m_\pi^2F_\pi ^2(1-\cos\theta )$ (and is different from that in \cite{taylor}).
The quasi-Goldstone modes excited from this $\theta$-vacuum are not generally
stable. To see this, we shift the fields as
\begin{eqnarray}
\phi\longrightarrow e^{i\theta (\vec{n}\cdot \vec{\tau})} (F_\pi +\phi)\quad ;
\quad
\phi^\dagger\longrightarrow
(F_\pi +\phi^\dagger)e^{-i\theta (\vec{n}\cdot \vec{\tau})}
\end{eqnarray}
and the would-be Goldstone bosons have a $(\mbox{mass})^2$
\begin{eqnarray}
\tilde{m}_\pi^2\simeq \frac{c}{F_\pi}\cos\theta=m_\pi^2\cos\theta\; .
\label{c}
\end{eqnarray}
(\ref{c}) can also be calculated by $\tilde{m}_\pi^2=\frac{1}{F_\pi}
 \frac{\partial^2 E}{\partial \theta^2}$ \cite{dashen}. When
$\frac{\pi}{2}<\theta <
 \frac{3\pi}{2}$, one has negative $(\mbox{mass})^2$ modes. These modes are not
stable and
 long wavelength fluctuations will grow so that the interior with a vacuum
orientation $\theta$
 in the interval $(\frac{\pi}{2},\frac{3\pi}{2})$ may be a correlated region
and
 the pion production will be very structural. The energy density $E_\theta$ in
 this region is positive, which offers a good probability to observe coherent
 pion emissions.

It is potentially important to ask what happens to the $U(1)_A$ phase in the
disoriented chiral condensate since the vacua may also differ in
their $U(1)_A$ orientations. This query immediately runs into the question of
the role of the
axial anomaly in QCD. The reason why we have an approximate $SU(2)$-degeneracy
of the vacua is that the ISB is small compared to $\Lambda
_{QCD}$. However,
 the $U(1)_A$ symmetry is badly broken by the chiral anomaly or the
 instanton effects at zero temperature. Nevertheless, at high temperature, the
symmetry breaking caused by the
 anomaly is much softer than the ISB (the quark mass term).
 Because of the color screening
 effect, a typical instanton amplitude is generally suppressed at a high $T$
 \cite{gross}, e.g.,
\begin{eqnarray}
K(T)\propto \Lambda_{QCD}^2\left ( \frac{8\pi^2}{g^2(T)}\right )^{2N_c}e^{-
\frac{8\pi^2}{g^2(T)}}\sim \Lambda_{QCD}^2\left (\frac{\Lambda_{QCD}}{\pi
T}\right )^8
\left [ 9\ln \left(\frac{\pi T}{\Lambda_{QCD}}\right )\right ]^6
\end{eqnarray}
which provides a suppressing factor of $10^{-7}$ for a typical hadronic energy
scale
 if $T=2\Lambda_{QCD}$.
Therefore, if the chiral anomaly is indeed mediated by the instanton effects,
one has an additional approximate vacuum degeneracy in the $U(1)_A$ directions
 at high enough
temperatures which may be attainable in high energy collisions.
As a result, we have another freedom to
choose the vacuum orientation (the freedom is again slightly violated by the
mass term) if the instanton effects is effectively turned off at a high $T$. In
terms of the quark language, the quark condensate may be tilted into one of the
$U(1)_A$ directions
\begin{eqnarray}
\left\langle \bar q_Lq_R\right\rangle =-v^3e^{i\varphi}\quad\quad ;\quad\quad
\left\langle \bar q_Rq_L\right\rangle =-v^3e^{-i\varphi} \; .\label{d}
\end{eqnarray}

There are two points to be clarified. First, if the cooling is in equilibrium,
when the temperature drops from high $T$ to $T_c$, the $U(1)_A$-degeneracy is
completely lifted since the instanton effects become substantial at $T_c$ (one
may avoid this by arguing that instanton interactions only become strong at an
even
lower temperature). Thus an out-of-equilibrium condition is as necessary as in
the case of the $SU(2)$ disorientation. Indeed, if there is a quench, the VEV's
will stay around zero for some time as suggested by Rajagopal and Wilczek (RW)
\cite{rw}, a disoriented $U(1)_A$ phase is possible. Second, a
negative $(\mbox{mass})^2$ for the $U(1)_A$ particle (the $\eta$
particle if we restrict ourselves
to the two-flavor case) may not arise from the RW mechanism in the intermediate
stages as it is the case for pions. This is because $\eta$ is not a would-be
Goldstone boson, its $(\mbox{mass})^2$ is given
\begin{eqnarray}
m_{\eta}^2\simeq -\mu ^2(T)+2\lambda\langle \phi\rangle ^2+K(T).
\end{eqnarray}
When $T$ suddenly drops to zero, $K(0)$ is large and may yield a positive
$m^2_\eta$ if $K(0)> \mu^2(0)$. Although it is not very clear whether one needs
a large size of the correlated region in order to have the $CP$-violating
effects to be discussed below, it is still possible to have a negative $\eta$
$(\mbox{mass})^2$ from the other mechanism which we have pointed out earlier.

What might be the effects associated with the $U(1)_A$ phase disorientation in
high
energy collisions? There will be an anomalous probability distribution of
$\eta$
and pion productions if the interior is also correlated in the $U(1)_A$ phase
through the vacuum misalignment mechanism.
It would be very hard to measure such an effect since the
$\eta$ has a large mass.  However, there may be another possibility, that is,
the $CP$ violation in a disoriented $U(1)_A$ phase. The basic idea is
following.
Normally, if $CP$ is conserved in QCD, the chiral condensate in (\ref{d}) must
be real because of the vacuum alignment. If, under some extreme condition, the
vacuum condensate is misaligned with the ISB and the
chiral anomaly, it is on longer possible to define definite
$CP$ transformation properties (parities) of $\eta$ and pions that
are consistent in both the vacuum and the lagrangian. In general, these
 mesons whose parities would be
negative
are mixtures of $CP$-even and -odd states. As a result of the mixing, an
initially $CP$-odd state can decay to a $CP$-even state (e.g.\ $\eta\rightarrow
2\pi$) and $CP$ symmetry is violated.
This is what may happen in a rapid cooling process.
At a  high $T$ the vacua
are almost degenerate in their $U(1)_A$ phases, any $\varphi$-direction is
equivalent
to any other directions because $K(T)\sim 0$. There is no $CP$ violation even
if
$\varphi \neq 0,\pi$. Following a quench, the instanton interactions are
suddenly turned on and $K(0)\neq 0$, the misaligned configurations ($\varphi
\neq 0,\pi$) are not stable and can only relax to $\varphi =0$ through $CP$
violating processes, for example, $\eta\rightarrow 2\pi$ decays. Luckily, we
may
be able to observe these events in high energy collisions even though QCD is a
$CP$ conserving theory under a normal condition.

The theoretical description of the picture can be idealized by considering an
extended linear $\sigma$-model where $\phi$ representing $\bar q_Lq_R$ has
eight
degrees of freedom if only two flavors are relevant
\begin{eqnarray}
\phi \equiv \frac{1}{2} (\sigma +i\eta )+\frac{1}{2} (\vec{\alpha}+i\vec{\pi})
\cdot \vec{\tau}\; .
\end{eqnarray}
The potential for meson fields at zero temperature reads
\begin{eqnarray}
V(\phi^\dagger,\phi)=V_0(\phi^\dagger\phi)+V_m(\phi^\dagger,\phi) +
V_K(\phi^\dagger,\phi)
\end{eqnarray}
where $V_0$ is invariant under $U(2)_L\times U(2)_R$
\begin{eqnarray}
V_0=-\mu^2 \mbox{Tr}\phi^\dagger\phi
+\frac{1}{2} (\lambda_1-\lambda_2)\left (
\mbox{Tr}\phi^\dagger\phi \right )^2 +\lambda_2  \mbox{Tr}\left (
\phi^\dagger\phi \right )^2\; ,
\end{eqnarray}
the ISB mass term
\begin{eqnarray}
V_m=-\frac{1}{4}c\left ( \mbox{Tr}\phi +\mbox{Tr}\phi^\dagger \right )\; ,
\end{eqnarray}
and the instanton-induced $U(1)_A$-breaking term \cite{huang}
\begin{eqnarray}
V_K=-K\mbox{det}\phi^\dagger -K\mbox{det}\phi\; .
\end{eqnarray}
We consider a $CP$ conserving theory by neglecting
 possible strong $CP$ phases in $V_m$ and $V_K$.
In this extended model, the disorientation in isospin
space is described by the non-zero VEV's of $\vec{\alpha}$ fields (rather than
$\langle\vec{\pi}\rangle\neq 0$). We will concentrate on the disorientation of
the $U(1)_A$ phase. The VEV of $\phi$ as a relic of high temperature history is
assumed
\begin{eqnarray}
\langle\phi\rangle=F_\pi e^{i\varphi}\; .
\end{eqnarray}
The dynamics of the excited fields from this vacuum is obtained by shifting the
fields (to get rid of tadpole terms)
$\phi\rightarrow e^{i\varphi}(F_\pi +\phi)$. The vacuum energy density is
calculated in the lowest order
\begin{eqnarray}
E_\varphi =-\frac{1}{4} KF_\pi^2 \cos 2\varphi -cF_\pi\cos\varphi
\end{eqnarray}
which is minimized at $\varphi =0$. Thus a disoriented vacuum has a higher
energy density. The $\eta$ $(\mbox{mass})^2$ reads from the shifted
potential or simply from
$\frac{1}{F_\pi^2}\frac{\partial^2E}{\partial\varphi ^2}$
\begin{eqnarray}
\tilde{m}_\eta^2=K\cos 2\varphi +m_\pi^2\cos\varphi\; .
\end{eqnarray}
Clearly $\eta$ acquires an additional mass from $V_K$. Like $\tilde{m}_\pi^2$
in
(\ref{c}), $\tilde{m}_\eta^2$ can be negative. The formation of a large size of
$\varphi$-oriented region is also possible. The energy density
$E_\varphi$ is associated with the disorientation of the $U(1)_A$ phase; it can
be released by $CP$ violating processes. The $CP$ violation occurs because it
is
not possible to define $CP$ transformation property of $\phi$; in particular,
$\sigma$ ($\vec{\alpha}$) will mix with $\eta$ ($\vec{\pi}$) in the presence of
$V_K$ via an interaction
\begin{eqnarray}
K\sin 2\varphi (\sigma\eta -\vec{\alpha}\cdot \vec{\pi})\; .
\label{x}
\end{eqnarray}
Such a mixing is genuine only if the $U(1)_A$-breaking
term $V_K$ is present. Equivalently, one can diagonalize (\ref{x}) by
redefining the physical states, and $CP$ violating vertices can be read off
from
the potential. For example, the amplitude of $\eta\rightarrow 2\pi$ decays
reads ($K=m_\eta^2-m_\pi^2$)
\begin{eqnarray}
A(\eta\rightarrow 2\pi )=\frac{1}{4}\frac{m_\eta^2-m_\pi^2}{F_\pi}
\sin 2\varphi
\end{eqnarray}
and the decay rate is estimated
\begin{eqnarray}
\Gamma (\eta\rightarrow 2\pi )=\frac{2}{256\pi}\frac{1}{m_\eta}
\left (\frac{m_\eta^2-m_\pi^2}{F_\pi}\right )^2
\sin^2 2\varphi
\sim 165\sin^2 \varphi\cos^2\varphi \;\mbox{MeV}.
\end{eqnarray}
This is expected because if the $\pi\pi$ channels are allowed, they become the
most dominant decay modes. It is very important to note that unlike a normal
$\eta$ particle whose decay rate is about a kilovolt, the $\eta$ in the DCC can
decay very fast ($\sim$ several fm's) so that it could be in a full contact
with the condensate before it decays and
we may be able to
observe such decay modes.

There are some theoretical and experimental uncertainties in this picture. Do
we need a large size of $U(1)_A$-phase correlated region to observe
$\eta\rightarrow
2\pi$ decays? We suspect so because otherwise we would average the amplitude
quantum mechanically
\begin{eqnarray}
\frac{1}{2\pi}\int A(\eta\rightarrow 2\pi )d\varphi =0\; .\label{f}
\end{eqnarray}
In this case, the negativity of $\tilde{m}^2_\eta$ is very crucial to implement
the
idea. But it is not clear that (\ref{f}) is really legitimate if different
domains in
the interior are independent in the $U(1)_A$ directions. Another question is
how
many $\eta$'s can be produced in an event. Following \cite{taylor}, the total
energy in the interior (mod $E_0$) is
\begin{eqnarray}
E\simeq \pi R^3m^2_\eta F_\pi^2 (1-\cos 2\varphi )\simeq 28 (1-\cos 2\varphi )
\;\mbox{GeV}
\end{eqnarray}
where $R\sim 3\mbox{fm}$. Thus there may be as many as  10 $\eta$-mesons.

In summary, we have studied some theoretical aspects in an interesting proposal
to observe the disoriented chiral condensate in high energy collisions. In
particular, we have suggested an additional mechanism to have a negative
$(\mbox{mass})^2$ for
the mesons responsible for the formation of a large size of the correlated
region. A possibility to observe unsuppressed strong $CP$ violation such as
$\eta\rightarrow 2\pi$ decays is pointed out. This study may shed some
light on the
issues of chiral anomaly, instanton effects including the color
screening mechanism and $CP$ violation in strong interactions.\\
\vspace{.2in}
\begin{center}
{\bf Acknowledgement}
\end {center}

I wish to thank J.\ Bjorken, M.\ Suzuki and Dandi Wu
for very useful discussions.
This work was supported by the Director, Office of Energy
Research, Office of High Energy and Nuclear Physics, Division of High
Energy Physics of the U.S. Department of Energy under Contract
DE-AC03-76SF00098 and by the
 Natural Sciences and Engineering
Research Council of Canada.
\pagebreak

\end{document}